# Transferred Energy Management Strategies for Hybrid Electric Vehicles Based on Driving Conditions Recognition

Teng Liu
Department of Automotive Engineering and the State Key Laboratory of Mechanical Transmission
Chongqing University, Chongqing 400044, China
tengliu17@gmail.com

Xiaolin Tang
Department of Automotive Engineering and the State Key Laboratory of Mechanical Transmission
Chongqing University, Chongqing 400044, China
tangxl0923@cqu.edu.cn

Jiaxin Chen
Department of Automotive Engineering
Chongqing University, Chongqing 400044, China
201932132050@cqu.edu.cn

Hong Wang
School of Vehicle and Mobility
Tsinghua University, Beijing 100084, China
hong_wang@tsinghua.edu.cn

Wenhao Tan
Department of Automotive Engineering
Chongqing University, Chongqing 400044, China
2471591789@qq.com

Yalian Yang
Department of Automotive Engineering and the State Key Laboratory of Mechanical Transmission
Chongqing University, Chongqing 400044, China
YYL@cqu.edu.cn

*Abstract*—Energy management strategies (EMSs) are the most significant components in hybrid electric vehicles (HEVs) because they decide the potential of energy conservation and emission reduction. This work presents a transferred EMS for a parallel HEV via combining the reinforcement learning method and driving conditions recognition. First, the Markov decision process (MDP) and the transition probability matrix are utilized to differentiate the driving conditions. Then, reinforcement learning algorithms are formulated to achieve power split controls, in which Q-tables are tuned by current driving situations. Finally, the proposed transferred framework is estimated and validated in a parallel hybrid topology. Its advantages in computational efficiency and fuel economy are summarized and proved.

## I. Introduction

Energy management strategies (EMSs) enable hybrid electric vehicles (HEVs) to achieve energy conservation and emission reduction [1-4]. By designing appropriate power split controls, the onboard internal combustion engine (ICE) and battery pack could provide power reasonably, in order to reduce fuel consumption and prolong battery lifespan [5, 6]. Therefore, academic and industrial communities seek to generate optimal energy management controls for different hybrid powertrains in various driving conditions [7, 8].

Many technologies have been applied in HEVs' energy management field, and they are typically classified as rule-based and optimization-based methods. For example, dynamic programming (DP) [9], convex programming (CP) [10], fuzzy logic control [11], Pontryagin's minimum principle (PMP) [12] and so on are leveraged to obtain the globally optimal controls for academic research purpose. As alternatives, occupying the advantages of high robustness and low calculative requirement, current commercial hybrid vehicles almost choose rule-based control strategies [13, 14]. The difficulty locates in that the yielding energy management controls should adapt to current driving conditions and be able to transform with the driving environments.

In recent years, artificial intelligence approaches play a critical role in HEV's energy management problem. Deep learning (DL) and reinforcement learning (RL) are the most popular choices due to their independence of powertrain configuration and adaptation of driving situations. For example, the authors in [15] presented two methods to predict the future driving cycles, by combining with the RL framework, two resulted predictive EMSs are generated and compared in this manuscript. Wu et al. focused on a deep deterministic policy gradients-based EMS for the hybrid bus over continuous spaces [16]. The relevant simulation results show that the new control strategy exhibits performance close to the optimal global DP. Ref. [17] constructed a bi-level control architecture to obtain the RL-based power split controls, wherein the higher-level discussed how to predict power demand in real-time and the lower-level used model-free RL algorithm to solve the related optimal control problems. However, the enormous training data and unexpected driving conditions restrict the performance of learning-based energy management controls.

To make the RL-based power split controls adaptive to different driving conditions, this work proposed a transferred control framework to merge driving conditions recognition and RL algorithm. First, the driving conditions are interpreted as driving cycles for HEVs, Markov decision process (MDP) is used to simulate the driving cycles, and transition probability matrix (TPM) are utilized to quantify the differences between multiple driving cycles. Then, Q-table in the RL algorithm is underlined to generate the power split controls for energy management problems. Thus, the differences in driving conditions result in the regulation of Q-table in the RL framework. Finally, the performance and efficiency of the proposed energy management controls are compared with the original ones, their advantages in computational efficiency and fuel economy are demonstrated and analyzed. The preferences and future development of RL-based EMSs in different driving situations are also specified and outlook.

The construction of the rest of the paper is organized as follows: how to recognize current driving conditions are discussed in Section II. The RL algorithm and adjustment of Q-table in the RL algorithm are described in Section III, wherein the elements of RL algorithms are shown in detail. Simulation results are evaluated and analyzed in Section IV, and Section V concludes the paper.

## II. Driving Conditions Recognition

In the content of this section, the approach for driving conditions recognition is illuminated. Moreover, the MDP for driving cycle and TPM for differences in driving cycles are introduced. By doing this, the driving situations of HEVs could easily be transformed, and the driving conditions are easily represented.

### A. Driving Cycle Modeling

Driving conditions for road vehicles mean vehicle speed, road grade, environment temperature, air density, pavement type, traffic information, and so on. In the energy management field of HEVs, driving conditions are mainly interpreted as vehicle speed, road grade, and air density. As road grade and air density are totally fixed, and they can be measured in advance, the velocity of wheels is the dominating parameter, which would influence the allocation of power among different energy sources. Therefore, the driving conditions represent driving cycles in this work, and they are the speed sequences that changed with the time horizon.

Recognizing current driving conditions in HEVs means recording the variation of vehicle speed. Since the powertrain specification is settled for a special vehicle, the power demand could be calculated based on the driving cycle information. Assuming a normal vehicle speed sequence is indicated as $V=\{v_i| i=1, …, N\} \in \mathbf{R}$, and it complies with Markov property, which signifies that the next velocity point is only decided by the previous one point. Based on this assumption, the driving cycle can be treated as the Markov chain (MC) or Markov decision process (MDP), and the transition from one-speed point to another is easily counted. For a particular driving cycle, the transition probability from current speed point to the next future one is computed as

$$\pi_{ij} = P(v^{+1} = v_j | v = v_i) = \frac{K_{ij}}{K_{ix}} \quad (1)$$

where $v^{+1}$ and $v$ are the next one-step ahead and present speed points, $\pi_{ij}$ is the transition probability from $v_i$ to $v_j$, and $K_{ij}$ is the number of transition times from $v_i$ to $v_j$, $K_{ix}$ is the overall transition times initiated from $v_i$. Naturally, the following equation is satisfied

$$K_{ix} = \sum_{j=1}^{N} K_{ij} \quad (2)$$

In practical application, the velocity interval is separated into finite disjoint points $v_i \in [V_{min}: \triangle V: V_{max}]$, wherein $\triangle V$ is a certain speed gap, $V_{min}$ and $V_{max}$ are the lower and upper bounds. After the driving cycle is given, the nearest neighbor approach is capable of counting the arbitrary speed point, which means the random speed point belongs to the nearest disjoint coordinate. For all the pairs $(i, j)$, the transition probability between them is efficiently calculated and stored, and then these elements could constitute the transition probability matrix (TPM) $\Pi$. Finally, the one-step ahead velocity vector can be determined by the following matrix manipulation as

$$V^{+1} = V \cdot \Pi \quad (3)$$

where $V^{+1}$ and $V$ are the vector (velocity sequence) of vehicle speed.

### B. Quantization of TPM

For different driving cycles, their relevant TPMs could be obtained by using the MC modeling. Apparently, the TPMs are diametrically peculiar for two different driving conditions (driving cycles). Hence, TPMs are able to be used for quantization and recognition of the driving condition. As the essence of TPM is a matrix, and thus the induced matrix norm (IMN) [7] is applied to quantify the differences between two TPMs.

Since $\Pi_1$ and $\Pi_2$ ($N \times N$ matrix) are two TPMs related to disparate driving cycles, the IMN measurement is defined as follows

$$IMN(\Pi_1 \| \Pi_2) = \| \Pi_1 - \Pi_2 \|_2 = \sup_{P \in R^N} \frac{|(\Pi_1 - \Pi_2)P|}{|P|} \quad (4)$$

where $\|\Pi\|_2$ indicates the second-order norm of matrix $\Pi$. Sup means the supremum of a scalar, and $P$ is an arbitrary non-zero vector, and it contains transition probability elements. Furthermore, to improve the computational efficiency, the second-order norm is represented by the maximum eigenvalue as following

$$IMN(\Pi_1 \| \Pi_2) = \max_{1 \leq j \leq N} \sqrt{\lambda_j[(\Pi_1 - \Pi_2)^T(\Pi_1 - \Pi_2)]} \quad (5)$$

where $\lambda_j(\Pi)$ indicates the $j$-th eigenvalue of matrix $\Pi$, $1 \leq j \leq N$. It is obvious that the IMN measurement is a scalar, and it can differentiate two TPMs. The real-time recognition process of driving conditions for energy management problems of HEVs is sketched in Fig. 1. After doing this, an arbitrary fragment of the driving cycle should be recognized, and the differences between two random driving cycles are easily computed and represented in real-time.

### C. Powertrain Specification

The energy management problem in this work locates as an optimization control problem with specific state variables and control actions. The research object is a parallel topology, its configuration is illustrated in Fig. 2. The maximum torque of ICE is 900 Nm, and its rated power is 155 kW. The rated capacity and nominal voltage of the battery pack are 60 Ah and 312.5 V, respectively. The maximum speed, power and torque of the electric motor is 2400 rpm, 90 kW and 600 Nm. The essential parameters of this parallel topology are described in Table I [18].

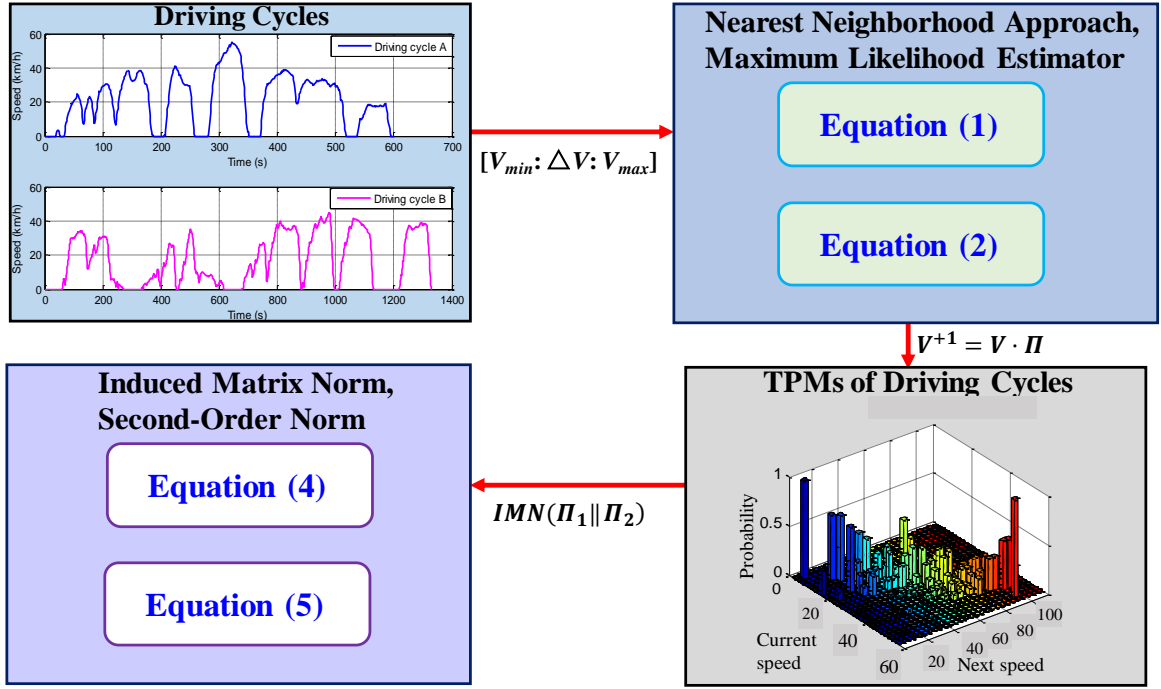

Fig. 1. Computational diagram of the driving conditions recognition.

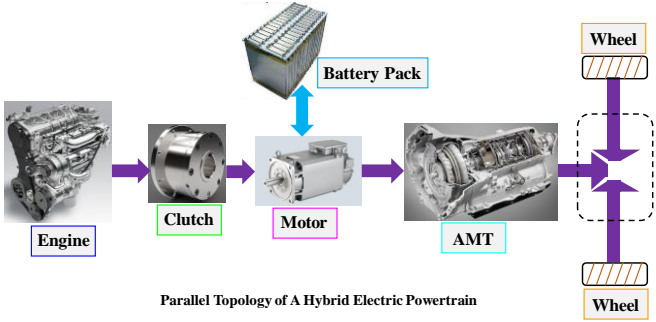

Fig. 2. Parallel hybrid configuration for energy management problem.

TABLE I
PARAMETERS OF MAIN COMPONENTS IN HEV

| Symbol | Items | Values |
| --- | --- | --- |
| $m$ | Vehicle mass | 16000 kg |
| $A$ | Fronted area | 1.8 m² |
| $C_d$ | Aerodynamic coefficient | 0.55 |
| $\eta_T$ | Transmission axle efficiency | 0.9 |
| $\eta_{mot}$ | Efficiency of Traction motor | 0.95 |
| $f$ | Efficiency of Rolling resistance | 0.021 |
| $R$ | Radius of Tire | 0.508 m |
| $\rho_a$ | Air density | 1.293 kg/m³ |
| $g$ | Gravitational acceleration | 9.81 m/s² |

In the optimization control (energy management) problem, the expected optimization objective is the systemization of charge sustenance and fuel economy. Thus, the cost function over a finite time horizon is given as

$$J = \int_{t_0}^{t_f} [\alpha(SOC(t) - SOC_{ref})^2 + m_f(t)]dt \quad (6)$$

where $[t_0, t_f]$ is the time interval of the driving cycle. SOC means the state-of-charge of the battery pack, which reflects the remaining electricity of battery. $\alpha$ is a positive weighting factor to restrict the terminal value of SOC ($\alpha$=10000 in this work), and $SOC_{ref}$ represents a pre-defined argument to guarantee the charge-sustaining constraint. $m_f$ indicates the instant fuel consumption rate of ICE, and it is given by a quasi-static model as follows

$$\begin{cases} m_f = f(T_e, \omega_e) \\ Fuel_{total} = \int_{t_0}^{t_f} m_f(t)dt \end{cases} \quad (7)$$

where $f$ is usually a lookup table function, and $T_e$ and $\omega_e$ are the torque and rotation speed of ICE. In this article, the ICE rotation speed $\omega_e$ is decided by the vehicle speed and transmission ratio. The ICE torque $T_e$ is defined as the control action in this optimization control problem. $Fuel_{total}$ means the cumulative fuel consumption over a given driving cycle.

Furthermore, SOC is selected as the state variable, and it is decided by a first-order internal resistance modeling, wherein $I_{bat}$ and $Q_{bat}$ indicate the electric current and rated capacity of the battery pack as

$$\begin{aligned} \dot{SOC} &= -I_{bat}(t)/Q_{bat} \\ \Rightarrow \dot{SOC} &= -(V_{oc} - \sqrt{V_{oc}^2 - 4r_{in}P_{bat}})/2r_{in}Q_{bat} \end{aligned} \quad (8)$$

where $V_{oc}$ is the open-circuit voltage of the battery pack, $r_{in}$ shows the internal resistance in the battery and $P_{bat}$ denotes the output power of the battery. Overall, the power demand of the hybrid powertrain is supplied by the ICE and battery pack, and thus the battery power can be calculated by the power demand and ICE power. Finally, the power demand is determined by the given driving cycle and powertrain specification, and the ICE power is decided by the control action and its rotation speed.

### III. TRANSFERRED RL ALGORITHMS

The transferred RL calculative framework is introduced in this section. First, the original Sarsa algorithm is constructed, including the state-action pair, reward function, and transition matrix. Then, the Q-table is underlined, and its transferred form is established by adding the IMN of two driving cycles. Finally, the essential factors of the RL algorithm and the transferred Sarsa algorithm are summarized.

*A. Quintuple Elements of RL*

In the RL framework, an intelligent agent learns to improve the performance of its action by communicating with the surrounding environment [19]. Before learning and interacting, the information of the environment can be known or not, which leads to two categories of RL algorithms, model-based and model-free ones [20]. In model-based algorithms, the modeling of the environment should be found first, and then the agent can attempt to obtain the optimal control policy based on it. In model-free algorithms, the modeling of the environment is not necessary, and the agent may spend more trials to collect the information.

The biggest difference between RL and other machine learning methods is that current control action decides not only immediate feedback but also the future ones. Hence, the selection of control action should consider long-term cumulative feedback. To professionally speaking, the quintuple elements of RL can be written as $<S, A, P, R, \beta>$, wherein $s \in S$ and $a \in A$ are the state variable and control action, they are set of SOC and ICE torque in this work. $r \in R$ indicates the feedback from the environment to the agent, and it is evaluated by the cost function in (6). $\beta$ is named as a discount factor, which is utilized to balance the importance of immediate and future cumulative rewards (feedbacks). Finally, $p \in P$ means the transition model of the state variable, and it is able to be derived by the TPM in the energy management problem.

The objective of the intelligent agent in RL is to search a control sequence $q$ to maximize the accumulated rewards. Value function $V(s)$ expresses the discounted accumulated rewards as follows [21]

$$V(s) = E(\sum_{t=t_0}^{t_f} \beta^t r(s,a)) \quad (9)$$

where $(s, a)$ is named as state-action pair, which can uniquely determine the reward. For iterative goal, the value function can be reformulated as the sum of instant reward and future cumulative rewards as

$$V(s) = (r(s,a) + \beta \sum_{s^{+1} \in S} p_{s,s^{+1}} V(s^{+1})) \quad (10)$$

where $s^{+1}$ is the next state variable computed by the state equation (8), $p_{s,s+1}$ is the transition probability from $s$ to $s^{+1}$. After rewriting the form of the value function, the optimal control action could be generated by minimizing the value function as following [22]

$$q^*(s) = \arg\min_a (r(s,a) + \beta \sum_{s^{+1} \in S} p_{s,s^{+1}} V(s^{+1})) \quad (11)$$

For each state variable, the corresponding optimal control action can be obtained via (11), and thus the optimal EMS for parallel hybrid powertrain in a special driving cycle is acquired through RL iteration.

*B. Transferred Sarsa Framework*

In practical application, another value function related to the state-action pair $(s, a)$ is applied to achieve the iteration process. It is called action-value function $Q(s, a)$ and its normal and optimum formats are represented as

$$\begin{cases} Q(s,a) = r(s,a) + \beta \sum_{s^{+1} \in S} p_{s,s^{+1}} Q(s^{+1}, a^{+1}) \\ Q^*(s,a) = r(s,a) + \beta \sum_{s^{+1} \in S} p_{s,s^{+1}} \min_{a^{+1}} Q(s^{+1}, a^{+1}) \end{cases} \quad (12)$$

where the difference between $V(s)$ and $Q(s, a)$ is the information of current control action is known or not. From (11) and (12), it is apparent that the optimal control policy in RL is completely determined by the action-value function $Q(s, a)$, which is also named as Q-table in this article.

To simply speaking, the control action minimizing the Q-table is the optimal control action in the RL framework. The ε-greedy policy is frequently employed to choose the control action at each time step. It indicates the intelligent agent explores a random action with probability ε to increase the information of the environment, and exploit the best action in the Q-table $Q(s, a)$ until now with probability 1-ε. The updating criterion of Q-table in Sarsa is depicted as following [23]

$$Q(s,a) \leftarrow Q(s,a) + \gamma [r + \beta Q(s^{+1}, a^{+1}) - Q(s,a)] \quad (13)$$

where current and next control actions ($a$ and $a^{+1}$) are both selected by the ε-greedy policy. $\gamma \in [0, 1]$ is a learning rate to trade-off the old and new learned information.

Assuming $Q_1$ is a mature Q-table related to one driving cycle 1 and its TPM $\Pi_1$, the mature Q-table $Q_2$ for another driving cycle 2 can be calculated by combining the existing $Q_1$ and IMN as

$$Q_2(s,a) = Q_1(s,a) \cdot IMN(\Pi_1 || \Pi_2) \quad (14)$$

$$q^*(s) = \arg\min_a (Q_2(s,a)) \quad (15)$$

Since these two driving cycles (1 and 2) are known in a prior, the IMN is easily computed. Q-table $Q_1$ is learned from (13), another Q-table $Q_2$ could be generated by (14) efficiently. By doing this, the optimal control policy (energy management policy) for driving cycle 2 is capable of being decided in an efficient way, and it will save plenty of time to learn the information of Q-table. The pseudo-code of the transferred Sarsa algorithm is depicted in Table II.

The proposed RL algorithm (transferred Sarsa algorithm) is implemented in Matlab through MDP toolbox [24]. The arguments in this algorithm are determined after a series of trials, wherein the learning rate $\gamma$ and discount factor $\beta$ are 0.95 and 0.1, respectively. The probability ε is equal to $0.1*0.99^t$

TABLE II
PSEUDO-CODE OF THE TRANSFERRED SARSA ALGORITHM

| RL Algorithm: Transferred Sarsa |
|---|
| 1. Initialize $Q_1(s, a)$, $s$, and two driving cycles |
| 2. Compute $\Pi_1$ and $\Pi_2$ related to two driving cycles |
| 3. Generate IMN measurement $IMN(\Pi_1\|\Pi_2)$ |
| 4. Transform Q-table $Q_2=Q_1 * IMN(\Pi_1\|\Pi_2)$ |
| 5. Repeat each step $t$=1, 2, 3… |
| 6. Choose $a$, based on $Q_2(s, \cdot)$ ($\varepsilon$-greedy policy) |
| 7. Taking action $a$, observe $r(s, a)$, $s^{+1}$ |
| 8. Define $q^*(s)=\arg\min_a Q_2(s^{+1}, a)$ |
| 9. $s \leftarrow s^{+1}$ |
| 10. Until $s$ is terminal |

## IV. SIMULATION RESULTS AND ANALYZATION

The corresponding simulation results of the transferred Sarsa in energy management problem is discussed in this section. The results based on the original Sarsa is named as Primary ones (derived by (13)), and the results related to the transferred Sarsa is called Transferred one (derived by (14) and (15)). First, these two methods are compared in the state variable and fuel consumption for two real-time driving cycles. Moreover, to prove the universality, these two algorithms are also implemented on multiple standard driving conditions.

### A. Comparison of Primary and Transferred Algorithms

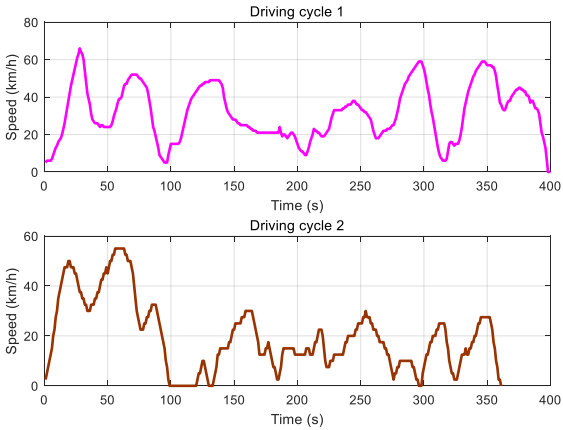

Fig. 3. Two driving cycles for comparison of primary and transferred control cases.

As shown in Fig. 3, two driving cycles (1 and 2) are used to evaluate the performance of primary and transferred Sarsa algorithms. Two control cases are carried out on the driving cycle 2, and the relevant effects are compared. In the primary case, the Q-table for driving cycle 1 is unknown in advance, so the agent should apply (13) to consummate the Q-table $Q_2$ and search the optimal EMS. In the transferred case, the Q-table for driving cycle 1 $Q_1$ is known, and thus the $Q_2$ and optimal EMS can be obtained by (14) and (15).

Fig. 4 depicts the SOC trajectories of these two compared control cases. It can be discerned that the SOC evolution is almost the same in these two situations. This is caused by the power split controls (or the energy management controls) between the ICE and battery pack. It indicates that the transferred Sarsa algorithm could generate nearly the same Q-table and optimal controls when compared with the primary one. Furthermore, this transferred algorithm could be finished in a more efficient way.

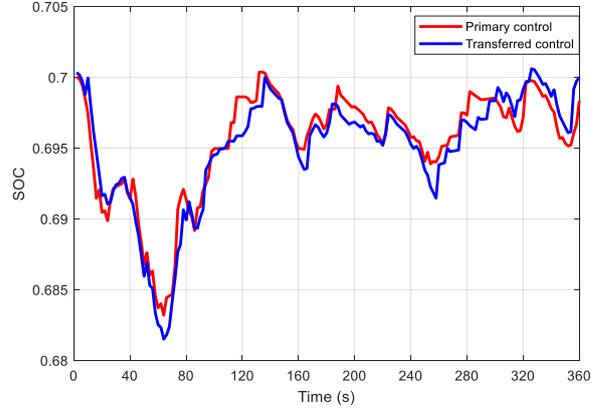

Fig. 4. SOC trajectories in primary and transferred control cases.

To compare the computational efficiency and fuel economy in these two methods, Table III describes the calculative time and fuel consumption of primary and transferred algorithms. It is noticed that the consumed fuel of the transferred algorithm is almost the same as that in the primary algorithm. Moreover, the consumed time in the transferred algorithm is much better than another one. Hence, it can be concluded that the proposed transferred control framework can achieve the same performance in a more efficient way.

TABLE III
TIME AND FUEL COSTS OF TWO CONTROL CASES

| Methods* | Time (s) | Fuel consumption (g) |
|---|---|---|
| Primary case | 86 | 450.62 |
| Transferred case | 37 | 437.39 |

\* A 2.90 GHz microprocessor with 7.83 GB RAM was used.

### B. Assessment of Adaptation

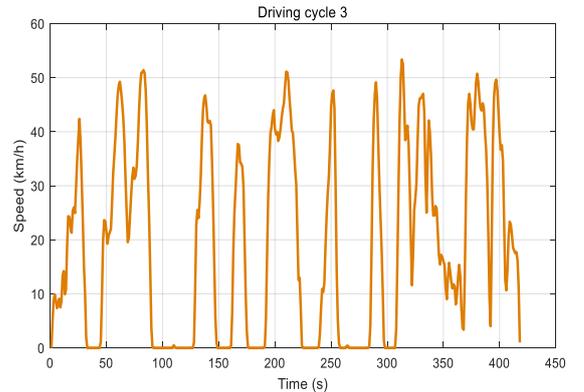

Fig. 5. Real-time driving cycle for evaluation of adaptability.

In this subsection, the presented transferred Sarsa algorithm is compared with the primary one and the benchmarking DP method on multiple standard driving cycles (including New European Driving Cycle (NEDC); Federal Test Procedure (FTP-75); Japanese Cycle (JC08)) to estimate its adaptability. Taking a real-time driving cycle 3 shown in

Fig. 5 as an example, the transferred Q-table would be implemented on this driving cycle. The state variable is SOC, control action is ICE torque, and the initial value of SOC is 0.7.

TABLE II
COMPARISON ANALYSIS OF CONSUMED FUEL IN DIFFERENT METHODS

| Algorithm* \ Cycles | NEDC | JC08 | FTP-75 |
|---|---|---|---|
| Primary control | 974.1& | 960.9 | 978.5 |
| Transferred control | 965.2 | 965.8 | 956.1 |
| DP | 936.5 | 936.5 | 936.5 |

\* A 2.90 GHz microprocessor with 7.83 GB RAM was used.
& Fuel consumption (g).

Table IV describes the relevant fuel cost in different control cases, in which the Q-tables learned from the standard driving cycles are known, and the transferred Q-table is applied to driving cycle 3. It is obvious that the control performance of primary and transferred controls in different cases is the same, and they are close to the results of DP. It implies that the proposed control framework is effective in different driving conditions, and it can save time to learn the Q-table. Moreover, since the consumed fuel is close to DP, the optimality of the presented transferred control framework can be guaranteed.

## V. CONCLUSION

A transferred RL control framework is formulated in this work to address the energy management problem of HEVs. The important element of energy management, driving condition, is treated as a driving cycle, and the online recognition method is proposed first. In the original Sarsa algorithm, a transferred Q-table is computed by combining the existing Q-table and IMN. IMN describes the differences between two driving cycles. Finally, the evaluation process indicates the transferred control framework can guarantee the optimality and adaptability. Future work focuses on extracting the essential elements from driving cycles and refining the transferred process in RL architecture.

ACKNOWLEDGMENT

To diagnose terminal lung cancer in August 2019, I have experienced a tough time from September 2019 to Feb. 2020. Hereby, I want to express my heartfelt thanks to all the people who show love and care for me. Especially, I want to thank my parents and my wife. I wish my two children could grow up healthily and happily. I wish people can recovery from COVID-19 as early as possible. Peace and Love!

REFERENCES

[1] T. Liu, X. Hu, W. Hu, and Y. Zou, "A heuristic planning reinforcement learning-based energy management for power-split plug-in hybrid electric vehicles," *IEEE Trans. Ind. Informat.*, vol. 15, no. 12, pp. 6436-6445, 2019.
[2] C. Martinez, X. Hu, D. Cao, and E. Velenis, "Energy management in plug-in hybrid electric vehicles: Recent progress and a connected vehicles perspective," *IEEE Trans. Veh. Technol.*, vol. 66, no. 6, pp. 4534-4549, June. 2017.
[3] T. Liu, X. Tang, H. Wang, H. Yu, and X. Hu, "Adaptive Hierarchical Energy Management Design for a Plug-In Hybrid Electric Vehicle," *IEEE Trans. Veh. Technol.*, vol. 68, no. 12, pp. 11513-11522, 2019.
[4] L. Serrao, S. Onori, and G. Rizzoni, "A comparative analysis of energy management strategies for hybrid electric vehicles," *J. Dyn. Sys. Meas. Control.*, vol. 133, no. 3, pp. 1-9, 2011.
[5] B. Chen, X. Li, S. Evangelou, and R. Lot, "Joint Propulsion and Cooling Energy Management of Hybrid Electric Vehicles by Optimal Control," *IEEE Trans. Veh. Technol.*, vol. 1, no. 99, 2020.
[6] Y. Zou, T. Liu, D. X. Liu, and F. C. Sun, "Reinforcement learning-based real-time energy management for a hybrid tracked vehicle," *Appl. Energy*, vol. 171, pp. 372-382, 2016.
[7] T. Liu, B. Wang, and C. Yang, "Online Markov Chain-based energy management for a hybrid tracked vehicle with speedy Q-learning," *Energy*, vol. 160, pp. 544-555, 2018.
[8] S. Uebel, N. Murgovski, C. Tempelhahn, and B. Baker, "Optimal energy management and velocity control of hybrid electric vehicles," *IEEE Trans. Veh. Technol.*, vol. 67, no. 1, pp. 327–337, 2018.
[9] A. Sierra, V. Herrera, A. Gonzalez-Garrido, A. Milo, H. Gaztanaga, H. Camblong, "Experimental comparison of energy management strategies for a hybrid electric bus in a test-bench," in *Proc. EVER*, pp. 1-9, 2018.
[10] X. Hu, Y. Zou, and Y. Yang, "Greener plug-in hybrid electric vehicles incorporating renewable energy and rapid system optimization," *Energy*, vol. 111, pp. 971–980, 2016.
[11] D. Zhou, and A. AI-Durra, et al., "Online energy management strategy of fuel cell hybrid electric vehicles based on data fusion approach," *Journal of Power Sources*, vol. 366, pp. 278-291, 2017.
[12] C. Sun, H. He, et al., "The role of velocity forecasting in adaptive-ecms for hybrid electric vehicles," in *Proc. The 7th International Conference on Applied Energy – ICAE2015*, vol. 75, pp. 1907-1912, 2015.
[13] I. B. Ali, M. Turki, J. Belhadj, and X. Roboam, "Optimized fuzzy rule-based energy management for a battery-less PV/wind-BWRO desalination system," *Energy*, vol. 159, pp. 216-228, 2018.
[14] Y. Wang, Z. Sun, and Z. Chen, "Rule-based energy management strategy of a lithium-ion battery, supercapacitor and PEM fuel cell system," *Energy Procedia*, vol. 158, pp. 2555-2560, 2019.
[15] T. Liu, Y. Zou, D. Liu and F. C. Sun, "Reinforcement learning of adaptive energy management with transition probability for a hybrid electric tracked vehicle," *IEEE Trans. Ind. Electron.*, vol. 62, no. 12, pp.7837-7846, 2015.
[16] Y. Wu, H. Tan, J. Peng, H. Zhang, and H. He, "Deep reinforcement learning of energy management with continuous control strategy and traffic information for a series-parallel plug-in hybrid electric bus," *Applied energy*, vol. 247, pp. 454-466, 2019.
[17] T. Liu, and X. Hu, "A Bi-Level Control for Energy Efficiency Improvement of a Hybrid Tracked Vehicle," *IEEE Trans. Ind. Informat.*, vol. 14, no. 4, pp. 1616-1625, 2018.
[18] T. Liu, X. Hu, S. Li, and D. Cao, "Reinforcement learning optimized look-ahead energy management of a parallel hybrid electric vehicle," *IEEE/ASME Trans. Mechatronics*, vol. 22, no. 4, pp. 1497-1507, 2017.
[19] R. Sutton, and A. Barto, "Reinforcement learning: An introduction (second edition)," *MIT press*, 2018.
[20] T. Liu, H. Yu, H. Guo, Y. Qin, and Y. Zou, "Online energy management for multimode plug-in hybrid electric vehicles," *IEEE Trans. Ind. Informat.*, vol. 15, no. 7, pp. 4352-4361, Nov 2018.
[21] K. Hwang, W. Jiang, and Y. Chen, "Pheromone-based planning strategies in Dyna-Q learning," *IEEE Trans. Ind. Informat.*, vol. 13, no. 2, pp. 424-435, 2017.
[22] T. Liu, Y. Zou, D. Liu and F. Sun, "Reinforcement learning-based energy management strategy for a hybrid electric tracked vehicle," *Energies*, vol. 8, no. 7, pp.7243-7260, 2015.
[23] M. Santos, J. Martin, V. Lopez, and G. Botella, "Dyna-*H*: A heuristic planning reinforcement learning algorithm applied to role-playing game strategy decision systems," *Knowledge-Based Systems*, vol. 32, pp. 28-36, 2012.